\documentclass[12pt]{iopart}

\usepackage{cite} 
\usepackage[colorlinks, linkcolor=blue, anchorcolor=blue, citecolor=blue, urlcolor=blue]{hyperref}
\usepackage{graphicx}
\usepackage{dcolumn}
\usepackage{bm}
\usepackage{multirow}
\usepackage{longtable}
\usepackage{supertabular}
\usepackage{booktabs}
\usepackage{threeparttable}
\usepackage{float}
\usepackage{subfigure}
\usepackage{caption2}
\usepackage{multirow}
\usepackage{amssymb}
\usepackage{graphicx}
\usepackage[normalem]{ulem}
\usepackage{array}
\usepackage{etoolbox}
\usepackage{setspace}
\usepackage{tabularx}
\usepackage{color}
\usepackage{cancel}

\begin{document}

\title{Half-lives for proton emission and $\mathcal{\alpha}$ decay within the deformed Gamow-like model}

\author{Qiong Xiao$^{1}$, Jun-Hao Cheng$^1$ and Tong-Pu Yu$^1$}

\address{$^1$Department of Physics, National University of Defense Technology, Changsha 410073, People's Republic of China}
\ead{\mailto{tongpu@nudt.edu.cn}}
\ead{\mailto{cjh452002@163.com}}

\vspace{10pt}
\begin{indented}
\item[]
\end{indented}

\begin{abstract}
In the present work, we study $\mathcal{\alpha}$ decay and proton emission half-lives within the modified Gamow-like model, which introduces the effects of the nucleus's deformation. The calculations show that it is necessary to consider the deformation in the calculation for nuclei far from the shell. Moreover, we use the improved model to predict the proton emission half-lives of the nuclei far from the shell. The calculation results indicate that our model is in good agreement with most models. Furthermore, the deformed Gamow-like model is used to find the following neutron magic number. This work is meaningful for future research on superheavy nuclei.
\end{abstract}

%
%
%
%
%

\section{Introduction}
\label{section 1}

The study of nuclear decay is an irreplaceable approach to understanding the structure of nuclei. $\mathcal{\alpha}$ decay and proton emission of nuclei were first discovered by Rutherford and Jackson in 1899 and 1970, respectively. Since the discovery of these two nuclear decay modes with the same physical processes, studying them theoretically and experimentally has always been a popular topic in nuclear physics. $\mathcal{\alpha}$ decay is an essential tool for studying superheavy nuclei and can provide valuable information about the nuclear structure and stability of superheavy nuclei \cite{Oganessian_2015}. The study of proton emission can extract meaningful information about the nuclear structure beyond the proton drip line, e.g., the shell structure, the coupling between unbound and bound nuclear states and so on \cite{KARNY200852}. Nowadays, there are many models used to study $\mathcal{\alpha}$ decay, such as the unified model for $\mathcal{\alpha}$ decay and $\mathcal{\alpha}$ capture \cite{PhysRevC.73.031301, PhysRevC.92.014602}, the empirical formulas \cite{PhysRevC.85.044608,0954-3899-39-1-015105, PhysRevC.80.024310, 0954-3899-42-5-055112}, the two-potential approach \cite{1674-1137-41-1-014102, PhysRevC.94.024338, PhysRevC.93.034316, PhysRevC.95.014319, PhysRevC.95.044303, PhysRevC.96.024318, PhysRevC.97.044322}, the cluster model \cite{PhysRevLett.65.2975, PhysRevC.74.014304, XU2005303}, the liquid drop model \cite{0305-4616-5-10-005, PhysRevC.48.2409, 0954-3899-26-8-305, PhysRevC.74.017304, GUO2015110} and others \cite{PhysRevLett.59.262, SANTHOSH201528, PhysRevC.87.024308,0954-3899-31-2-005, PhysRevC.81.064318, QI2014203}. There are also many theoretical models used to study proton emission or the different forms of interactions used to construct these models, such as the single-folding model \cite{PhysRevC.72.051601,Qian_2010}, the Coulomb and proximity potential model (CPPM) \cite{PhysRevC.96.034619}, the distorted-wave Born approximation \cite{PhysRevC.56.1762}, the R-matrix approach \cite{PhysRevC.85.011303}, the relativistic density functional theory \cite{FERREIRA2011508}, the generalized liquid-drop model \cite{PhysRevC.79.054330,Zhang_2010,PhysRevC.95.014302}, the phenomenological united fission model \cite{PhysRevC.71.014603}, the effective interactions of density-dependent M3Y (DDM3Y) \cite{BHATTACHARYA2007263,Qian_2010} and so on \cite{Chen_2019}. These models and methods reproduce, to varying degrees, experimental data on alpha decay and proton emission half-lives.

Recently, Zdeb \textit{et al.} \cite{PhysRevC.87.024308,1402-4896-2013-T154-014029} purported Gamow-like model, which has attracted much attention\cite{XING2022122528,Azeez2022,Liu_2021,CHENG2019350,Liu_2022,Chen_2019,Zhu2022}, as a simple phenomenological model from Gamow's theory that can be used to calculate the $\mathcal{\alpha}$-decay half-life. In this model, the nuclear potential is taken as a square potential well, the Coulomb potential is chosen as the potential of a uniformly charged sphere, and the centrifugal potential is neglected. Since proton emission has the exact physical mechanism as $\mathcal{\alpha}$-decay, they also extended the model for studying proton emission \cite{Zdeb2016}. The theoretical half-life obtained by Gamow-like model is sensitive to the position of the outer turning point. It is essential to improve the accuracy of the potential curve for the calculation. The nuclei capable of $\mathcal{\alpha}$ decay and proton emission cover a large number of parent nuclei far from the shell. Therefore, the deformation of parent nuclei can not be neglected in calculating $\mathcal{\alpha}$ decay and proton emission half-lives using Gamow-like model. In the present work, we modify Gamow-like model by considering the deformation of the nucleus to study $\mathcal{\alpha}$ decay and proton emission half-lives systematically. The calculations show that for parent nuclei far from the shell, the theoretical half-lives obtained by the improved model are in better agreement with the experimental data than Gamow-like model. Moreover, this property is applied to find the magic number of the next neutron.

This article is organized as follows. In Sec. II the theoretical framework for Gamow-like model considering the deformation is described in detail. In Sec. III, the detailed calculations, discussion and predictions are provided. A brief summary is given in Sec. IV.

\section{Theoretical framework}

In Gamow-like model, $\mathcal{\alpha}$ decay and proton emission half-lives can be given by the decay constant $\mathcal{\lambda}$
\begin{equation}
\label{1}
T_\frac{1}{2}=\frac{\rm{ln2}}{\lambda}10^h,
\end{equation}
where the hindrance factor $h$ is used to describe the influence of an odd-proton and/or an odd-neutron on $\mathcal{\alpha}$ decay. The decay constant $\mathcal{\lambda}$ can be expressed as \cite{PhysRevC.83.014601}
\begin{equation}
\label{2}
\lambda=S \nu P,
\end{equation}
where $S$ is the preformation probability of $\mathcal{\alpha}$ particle or proton on the surface of the nucleus. $\nu$ is the collision frequency of $\mathcal{\alpha}$ particle or proton in the potential barrier, which can be calculated by the oscillation frequency $\omega$ \cite{PhysRevC.81.064309}
\begin{equation}
\label{9}
\nu=\omega/2\pi=\frac{(2n_{r}+l+\frac{3}{2})\hbar}{2\pi \mu {R_{n}}^2}=\frac{(G+\frac{3}{2})\hbar}{1.2\pi \mu{R_{0}}^2},
\end{equation}
where $\hbar$ is the reduced Planck constant. $\mu=\frac{m_{d}m'}{(m_{d}+m')}$ represents the reduced mass of the daughter nucleus and the emitted particle ($\mathcal{\alpha}$ particle or proton) in the center-of-mass coordinate with $m'$ and $m_{d}$ being masses of the emitted particle and the daughter nucleus. The $R_{n}= \sqrt{3/5}R_{0}$ denotes the nucleus root-mean-square radius with the radius of the parent nucleus $R_0 = 1.28A^{1/3}-0.76+0.8A^{-1/3}$. And $A$ is the mass number of the parent nucleus. Moreover, $G=2n_{r}+l$ is the main quantum number \cite{PhysRevC.69.024614,Chen_2019} with $l$ and $n_{r}$ being the angular quantity quantum number and the radial quantum number, respectively. 

Based on the classical WKB (Wentzel–Kramers–Brillouin) approximation, the penetration probability $P$ in Gamow-like model can be written as
\begin{equation}
\label{3}
P=\exp\! [- \frac{2}{\hbar} \int_{R_{in}}^{R_{out}} \sqrt{2\mu (V(r)-E_{k})}\, dr],
\end{equation}
where $E_{k}={Q}{\frac{A-A'}{A}}$ represents the momentum of the emitted particle. $A'$ and $Q$ represent the proton number of the emitted particle and the decay energy, respectively. $V(r)$ is the total potential between the daughter nucleus and the emitted particle. In this framework, it can be given by
\begin{equation}
V(r)=\left\{
\begin{array}{rcl}
-V_0,& & {0 \leq r \leq R_{in},}\\
V_C(r)+V_l(r), & & {r \ge R_{in},}
\end{array} \right.
\end{equation}
where $V_0$ is the depth of the square potential well. $V_l(r)$ and $V_C(r)$ are the centrifugal potential and Coulomb potential, respectively. The radius of the spherical square well $R_{in}$ is the sum of the radii of both the emitted particle and daughter nucleus
\begin{equation}
\label{eq5}
R_{in}=r_0({A_{d}}^\frac{1}{3}+{A'}^\frac{1}{3}),
\end{equation}
where the radius constant $r_{0}$ is an adjustable parameter in this model. The outer turning point $R_{out}$ satisfies the condition $V(R_{out})=E_{k}$. $A_{d}$ represents the mass number of the daughter nucleus.

In Gamow-like model, the Coulomb potential describes the emitted particle--daughter nucleus electrostatic interactions, which is given by
\begin{equation}
\label{eq6}
V_C(r)=Z'Z_de^2/r,
\end{equation}
where $Z'$ and $Z_d$ are the proton number of the daughter nucleus and the emitted particle, respectively. The centrifugal potential $V_{l}(r)$ is written as the Langer modified form in this work due to $l(l + 1) \rightarrow (l + 1/2)^2$ is a necessary correction for one-dimensional problem \cite{Gur31}. It can be given by
\begin{equation}
\label{eq10}
V_{l}(r)=\frac{\hbar^2(l+\frac{1}{2})^2}{2{\mu}r^2}.
\end{equation}

By introducing the deformations, the radius of the square well $R_{in}(\theta)$ is rewritten as a function associated with the orientation angle of the emitted particle concerning the symmetry axis of the daughter nucleus $\theta $. It can be expressed as
\begin{equation}
R_{in}(\theta)=R_{in}(1+\beta_2 Y_{20}(\theta)+\beta_4 Y_{40}(\theta)+\beta_6 Y_{60}(\theta)),
\end{equation}
where $\beta_2$, $\beta_4$ and $\beta_6$ taken from FRDM2012 \cite{MOLLER20161} denote the calculated quadrupole, hexadecapole and hexacontatetrapole deformation of the nuclear ground-state, respectively. $Y_{ml}(\theta)$ is shperical harmonics function. Furthermore, the total penetration probability $P$ can be obtained by averaging over all directions of $P(r,\theta)$ \cite{PhysRevC.73.041301}

\begin{equation}
P=\frac{1}{2}\int_{0}^{\pi}P(r,\theta)\, sin\theta d\theta ,
\label{eq8}
\end{equation}
\begin{equation}
P(r,\theta)=\exp\! [- \frac{2}{\hbar} \int_{R_{in}(\theta)}^{R_{out}(\theta)} \sqrt{2\mu (V(r,\theta)-E_{k})}\, dr].
\label{eq9}
\end{equation}
where $V(r,\theta)$ is the total potential between the daughter nucleus and the emitted particle considering the deformation, which is rewritten as
\begin{equation}
V(r,\theta)=\left\{
\begin{array}{rcl}
-V_0,& & {0 \leq r \leq R_{in}(\theta),}\\
V_C(r,\theta)+V_l(r), & & {r \ge R_{in}(\theta),}
\end{array} \right.
\end{equation}

In the present work, the deformated Coulomb potential $V_C(r,\theta)$ is obtained by the double-folding model. It can be expressed as
\begin{equation}
V_C(\mathop{r}^{\rightarrow},\theta)=\int_{}^{}\int_{}^{}\frac{\rho_{1}(\vec{r_1})\rho_2(\vec{r_2})}{\mid \vec{r}+\vec{r_2}-\vec{r_1} \mid } d\vec{r_1}d\vec{r_2},
\end{equation}
where $\vec{r_1}$ and $\vec{r_2}$ are the radius vectors in the charge distributions of the daughter nuclei and the emitted particle, respectively. $\vec{r}$ is the vector between the centers of the emitted proton and daughter nucleus. $\rho_1$ and $\rho_2$ represent the density distribution of the daughter nucleus and the emitted particle, respectively. Simplified appropriately by Fourier transform \cite{PhysRevC.61.044607,ISMAIL200353,Gao_Long_2008}, the deformated Coulomb potential can be approximated as
\begin{equation}
V_C(\mathop{r}^{\rightarrow},\theta)=V_C^{(0)}(\vec{r},\theta)+V_C^{(1)}(\vec{r},\theta)+V_C^{(2)}(\vec{r},\theta),
\end{equation}
where $V_C^{(0)}(\vec{r},\theta)$, $V_C^{(1)}(\vec{r},\theta)$ and $V_C^{(2)}(\vec{r},\theta)$ represent the bare Coulomb interaction, linear Coulomb coupling and second-order Coulomb coupling, respectively \cite{PhysRevC.61.044607}.

\begin{figure}[hbt]
\centering 
\includegraphics[width=13.0cm]{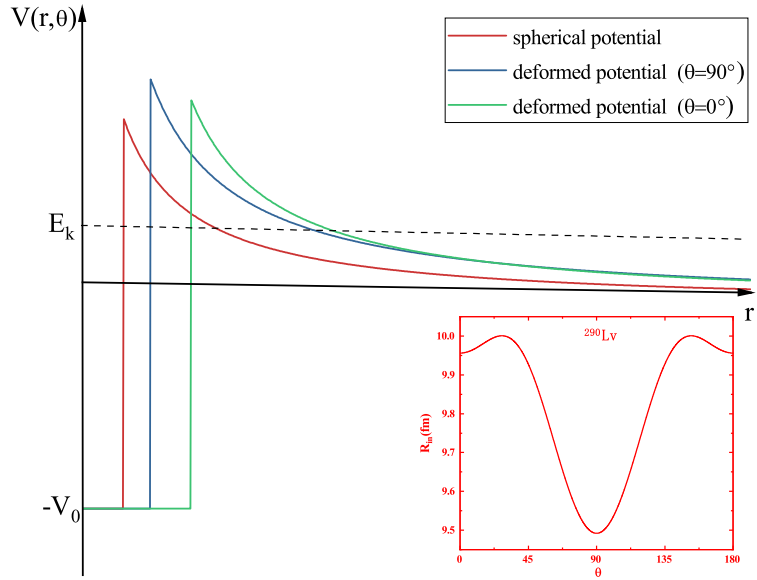}
\caption{Schematic diagram of the total potential between the daughter nucleus and the emitted particle with and without considering deformation.}
\label{fig 1}
\end{figure}

In this framework, the total potential curves between the daughter nucleus and the emitted particle with and without considering deformation are plotted in Figure \ref{fig 1}. In this figure, the red curve represents the total potential of the spherical approximation. The blue and green curves represent the orientation angle of the emitted particle concerning the symmetry axis of the daughter nucleus $\theta=90^{\circ}$ and $\theta=0^{\circ}$ corresponding to the deformed potential, respectively. The red box in the lower right corner depicts $R_{in}(\theta)$ of the parent nucleus $^{290}\rm{Lv}$ as $\theta$ changes. The figure shows that the introduction of deformation leads to a change in the total potential curve as well as the outer turning point.

\section{Results and discussion}
\label{section 3}
\subsection{proton emission}

In the present work, the least-squares principle redetermines the adjustable parameter $r_0=1.21 {\ }\rm{fm}$ of the deformed Gamow-like model for proton emission. The preformation probability of proton $S_p=1$ and the hindrance factor $h=0$ remain consistent with Ref.\cite{Zdeb2016}. For proton emission, the experimental data of half-lives, the decay energy, parity, and spin are taken from the latest evaluated atomic mass table AME2020 \cite{CPC-2021-0034,CPC-2020-0033} and the latest evaluated nuclear properties table NUBASE2020 \cite{NUBASE2020} except for those of $^{140}$Ho,$^{144}$Tm, $^{151}$Lu, $^{159}$Re, and $^{164}$Ir, which are taken from Ref. \cite{BLANK2008403}. By considering deformation, we have investigated the half-lives of proton emission for $51 \leq Z \leq 83$ nuclei. The detailed results are listed in Tab. \ref{tab1}. In this table, the first three columns are the parent nuclei, proton emission energy $Q_p$, and the orbital angular momentum $l$ taken away by the emitted proton, respectively. The following three columns are quadrupole $\beta_2$, hexadecapole $\beta_4$, and hexacontatetrapole $\beta_6$, respectively. The last three columns represent the logarithmic form of the experimental proton emission half-lives denoted as $lgT^{\rm{exp}}_{1/2}$, the logarithmic form of the theoretical proton emission half-lives calculated by Gamow-like model and the deformed Gamow-like model denoted as $lgT^{\rm{cal1}}_{1/2}$ and $lgT^{\rm{cal2}}_{1/2}$, respectively.

\begin{table}[H]
\caption{The calculations of the proton emission half-lives. $lgT^{\rm{exp}}_{1/2}$ represents the logarithmic form of the experimental proton emission half-lives. $lgT^{\rm{cal1}}_{1/2}$ and $lgT^{\rm{cal2}}_{1/2}$ are the theoretical proton emission half-lives calculated by Gamow-like model and the deformed Gamow-like model, respectively.}
\label{tab1}
\renewcommand\arraystretch{1.0}
\setlength{\tabcolsep}{2mm}
\begin{footnotesize}
\begin{tabular}{ccccccccc}
\hline
\hline
Nucleus &$l$&$Q_{p}(\rm{MeV})$ &$\beta_2$&$\beta_4$&$\beta_6$&$lgT^{\rm{exp}}_{1/2}$(s)& $lgT^{\rm{cal1}}_{1/2}$(s)& $lgT^{\rm{cal2}}_{1/2}$(s) \\
 \hline
$^{108}$I&0.597&2&0.15&0.071&-0.009&0.723&-0.056&0.196\\
$^{109}$I&0.82&0&0.162&0.06&-0.009&-4.032&-5.284&-5.052\\
$^{112}$Cs&0.816&2&0.196&0.054&-0.016&-3.31&-3.507&-3.266\\
$^{113}$Cs&0.973&2&0.207&0.056&-0.016&-4.771&-5.673&-5.437\\
$^{117}$La&0.82&2&0.282&0.106&0.005&-1.664&-2.852&-2.676\\
$^{121}$Pr&0.89&2&0.316&0.078&-0.014&-1.921&-3.222&-3.051\\
$^{131}$Eu&0.947&2&0.32&0.002&-0.016&-1.699&-2.706&-2.512\\
$^{135}$Tb&1.188&3&0.322&-0.037&-0.007&-2.996&-4.228&-4.035\\
$^{140}$Ho&1.106&3&0.289&-0.07&-0.002&-2.222&-2.722&-2.504\\
$^{141}$Ho&1.247&0&0.265&-0.062&0.002&-5.137&-5.933&-5.701\\
$^{141}$Ho&1.177&3&0.265&-0.062&0.002&-2.387&-3.557&-3.321\\
$^{144}$Tm&1.725&5&0.255&-0.076&0&-5.569&-5.48&-5.237\\
$^{145}$Tm$^m$&1.736&5&0.231&-0.068&0.004&-5.499&-5.564&-5.297\\
$^{146}$Tm&1.206&5&0.22&-0.069&0.005&-1.137&-1.087&-0.809\\
$^{146}$Tm&0.896&0&0.22&-0.069&0.005&-0.81&-0.691&-0.437\\
$^{147}$Tm&1.059&5&-0.187&-0.032&0.007&0.587&0.707&1.03\\
$^{147}$Tm$^m$&1.12&2&-0.187&-0.032&0.007&-3.444&-3.117&-2.834\\
$^{150}$Lu&1.29&2&-0.176&-0.045&0&-4.398&-4.419&-4.202\\
$^{150}$Lu&1.27&5&-0.176&-0.045&0&-1.347&-1.247&-0.987\\
$^{151}$Lu$^m$&1.301&2&-0.167&-0.035&0.007&-4.796&-4.545&-4.324\\
$^{151}$Lu$^m$&1.255&5&-0.167&-0.035&0.007&-0.896&-1.103&-0.837\\
$^{155}$Ta&1.453&5&0.021&0&0&-2.495&-2.543&-2.217\\
$^{156}$Ta$^m$&1.02&2&-0.073&0.002&0&-0.826&-0.505&-0.249\\
$^{156}$Ta&1.11&5&-0.073&0.002&0&0.933&1.18&1.497\\
$^{157}$Ta&0.935&0&0.085&0.003&-0.01&-0.527&0.044&0.281\\
$^{159}$Re&1.801&5&0.064&0.002&-0.01&-4.665&-4.776&-4.46\\
$^{159}$Re$^m$&1.816&5&0.064&0.002&-0.01&-4.678&-4.875&-4.559\\
$^{160}$Re&1.267&0&0.107&-0.008&0.009&-3.163&-3.845&-3.612\\
$^{161}$Re$^m$&1.317&5&0.118&0.005&0.01&-0.678&-0.744&-0.45\\
$^{161}$Re&1.197&0&0.118&0.005&0.01&-3.306&-3.046&-2.815\\
$^{164}$Ir&1.844&5&0.107&0.004&0.01&-3.959&-4.662&-4.362\\
$^{165}$Ir$^m$&1.711&5&0.118&0.005&0.01&-3.433&-3.747&-3.453\\
$^{166}$Ir&1.152&2&0.129&0.006&0&-0.824&-1.111&-0.864\\
$^{166}$Ir&1.332&5&0.129&0.006&0&-0.076&-0.398&-0.108\\
$^{167}$Ir$^m$&1.07&0&0.14&0.007&0&-1.12&-0.749&-0.518\\
$^{167}$Ir&1.245&5&0.14&0.007&0&0.842&0.57&0.852\\
$^{170}$Au$^m$&1.472&2&-0.105&-0.008&0.001&-3.487&-4.08&-3.822\\
$^{170}$Au&1.752&5&-0.105&-0.008&0.001&-3.975&-3.629&-3.32\\
$^{171}$Au$^m$&1.702&5&-0.115&-0.018&0.002&-2.587&-3.268&-2.963\\
$^{171}$Au&1.448&0&-0.115&-0.018&0.002&-4.652&-4.608&-4.368\\
$^{176}$Tl$^m$&1.265&0&0.075&-0.01&-0.001&-2.208&-2.102&-1.847\\
$^{177}$Tl$^m$&1.963&5&0.075&-0.01&-0.001&-3.346&-4.681&-4.359\\
$^{177}$Tl&1.172&0&0.075&-0.01&-0.001&-1.178&-0.926&-0.669\\
$^{185}$Bi$^m$&1.607&0&0.307&0.023&-0.009&-4.191&-5.052&-4.903\\
\hline
\hline
\end{tabular}
\end{footnotesize}
\end{table}

From Tab. \ref{tab1}, we can clearly see that for most parent nuclei, especially those far from the shell, the calculations obtained by the deformed Gamow-like model reproduce the experimental data better than Gamow-like model. The standard deviation $\sigma$ stands for the divergence between the experimental data and the theoretical proton emission half-lives, which is given by $\sigma=\sqrt{\sum ({\rm{lg}{T^{\rm{exp}}_{1/2}}(s)}-{\rm{lg}{T^{\rm{cal}}_{1/2}}(s)})^2/n}$. According to Tab. \ref{tab1}, we can conclude that the standard deviation ${\sigma_{\rm{cal1}}} = 0.618$, calculated by Gamow-like model, and the standard deviation ${\sigma_{\rm{cal2}}} = 0.556$, calculated by the deformed Gamow-like model. Therefore, considering the parent nucleus's deformation, the half-lives of the proton emission calculated in this work are within a factor of 3.59. For all parent nuclei, our calculations ${\sigma_{\rm{cal2}}}$ improve $(0.618-0.556)/0.618 \approx 10\%$ compared to ${\sigma_{\rm{cal2}}}$. 

\begin{figure}[hbt]
\centering 
\includegraphics[width=13.0cm]{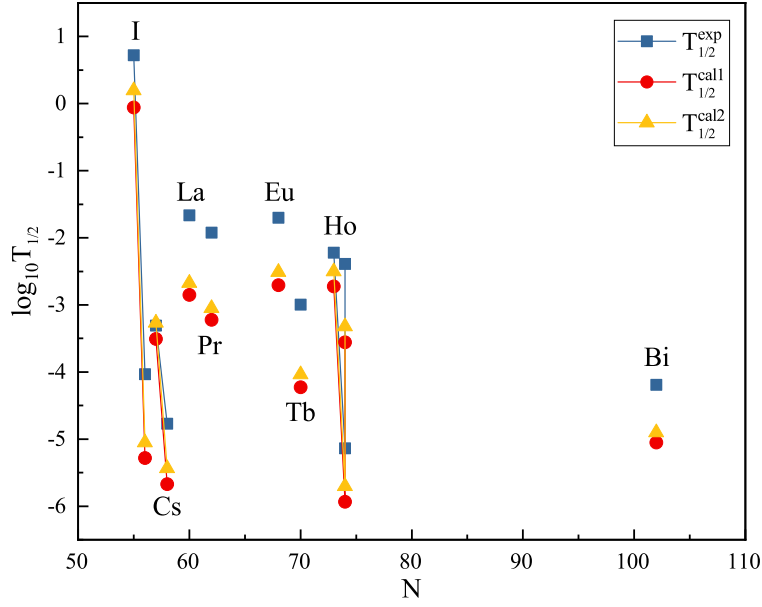}
\caption{The calculation of the proton emission half-life of the nuclei far from the shell. The blue squares stand for the experimental data, the red circles and yellow triangles are proton emission half-lives calculated by Gamow-like model and our work, respectively.}
\label{fig 2}
\end{figure}

In particular, for the 12 nuclei far from the shell ($Z \leq 67$ or $Z=83$), our standard deviation calculations significantly improved by 22\%. To show more visually the role of deformation in describing the half-lives of nuclei far from the shell, the experimental data and theoretical half-lives of these ones from Tab. \ref{tab1} are plotted in Fig. \ref{fig 2}. The X-axis represents the neutron number of the parent nucleus, and the Y-axis represents the logarithmic form of the half-lives of proton emission. The blue squares stand for the experimental data, the red circles and yellow triangles are proton emission half-lives calculated by Gamow-like model and our work, respectively. Fig. \ref{fig 2} shows that proton emission half-lives of the nuclei far from the shell vary over a wide range from $10^{-6}$ s to 10 s. Although the variation range of deformed proton emission half-lives is up to seven orders of magnitude, all the theoretical proton emission half-lives obtained by the deformed Gamow-like model are in better agreement with the experimental data than Gamow-like model. This implies that it is necessary to consider the deformation of the parent nucleus in Gamow-like model for the nuclei capable of emitting protons far from the shell.

Since the success of the deformed Gamow-like model in calculating the proton emission half-lives of the deformed nuclei, as an application, this model is used to predict the proton emission half-lives of five nuclei far from the shell, which is not yet quantified but observed in NUBASE2020 \cite{NUBASE2020}. The experimental data used for the predictions are all taken from NUBASE2020 \cite{NUBASE2020}, AME2020 \cite{CPC-2021-0034,CPC-2020-0033}, and FRDM2012 \cite{MOLLER20161}. As comparison, we also calculate the theoretical proton emission half-lives of these five nuclei using the universal decay law for proton emission (UDLP) \cite{PhysRevLett.103.072501,PhysRevC.80.044326,PhysRevC.85.011303}, Coulomb potential and proximity potential model with Guo-2013 formalism (CPPM-Guo2013) \cite{PhysRevC.93.024612,GUO201354,Deng2019}, two-potential approach with folding potentials model within density-dependent M3Y (TPA-DDM3Y) \cite{Chen2018}, the new Geiger–Nuttall law (NGNL) \cite{epjaChen2019}, and the two-potential approach based on Skyrme–Hartree–Fock within MQSP (TPA-SHF-MQSP) \cite{Cheng_2022}, respectively. The predicted results are given in Tab. \ref{tab2}. In this table, the first two columns are the parent nuclei, proton emission energy $Q_p$, and the following columns represent the theoretical proton emission half-lives calculated by UDLP, CPPM-Guo2013, TPA-DDM3Y, NGNL, TPA-SHF-MQSP, and our work denoted as ${\rm{lg}{T^{\rm{UDLP}}_{1/2}}}$, ${\rm{lg}{T^{\rm{CPPM}}_{1/2}}}$, ${\rm{lg}{T^{\rm{DDM3Y}}_{1/2}}}$, ${\rm{lg}{T^{\rm{NGNL}}_{1/2}}}$, ${\rm{lg}{T^{\rm{MQSP}}_{1/2}}}$ and ${\rm{lg}{T^{\rm{This-work}}_{1/2}}}$, respectively.

\begin{table*}[hbt!]
\caption{The predicted half-lives of deformed proton emission obtained by different models calculate.}
\label{tab2}
\renewcommand\arraystretch{1.1}
\setlength{\tabcolsep}{2mm}
\begin{footnotesize}
\begin{tabular}{cccccccc}
\hline
\hline
Nucleus &$Q_{p}(\rm{MeV})$ &${\rm{lg}{T^{\rm{UDLP}}_{1/2}}}$(s)&${\rm{lg}{T^{\rm{CPPM}}_{1/2}}}$(s)&${\rm{lg}{T^{\rm{DDM3Y}}_{1/2}}}$(s)&${\rm{lg}{T^{\rm{NGNL}}_{1/2}}}$(s)& ${\rm{lg}{T^{\rm{MQSP}}_{1/2}}}$(s)& ${\rm{lg}{T^{\rm{This-work}}_{1/2}}}$(s) \\
 \hline
$^{111}$Cs&1.731&-9.862&-11.252&-11.431&-10.101&-11.687&-11.258\\
$^{127}$Pm&0.781&-0.099&-0.166&-0.481&0.286&-0.751&-0.53\\
$^{137}$Tb&0.831&0.337&0.356&-0.031&0.739&-0.197&0.065\\
$^{185}$Bi&1.523&-0.67&-0.61&-0.637&-0.35&-0.852&-0.818\\
$^{185}$Bin$^n$&1.703&-0.859&-0.92&-0.861&-1.163&-1.036&-1.169\\
\hline
\hline
\end{tabular}
\end{footnotesize}
\end{table*}

\begin{figure}[hbt]
\centering 
\includegraphics[width=13.0cm]{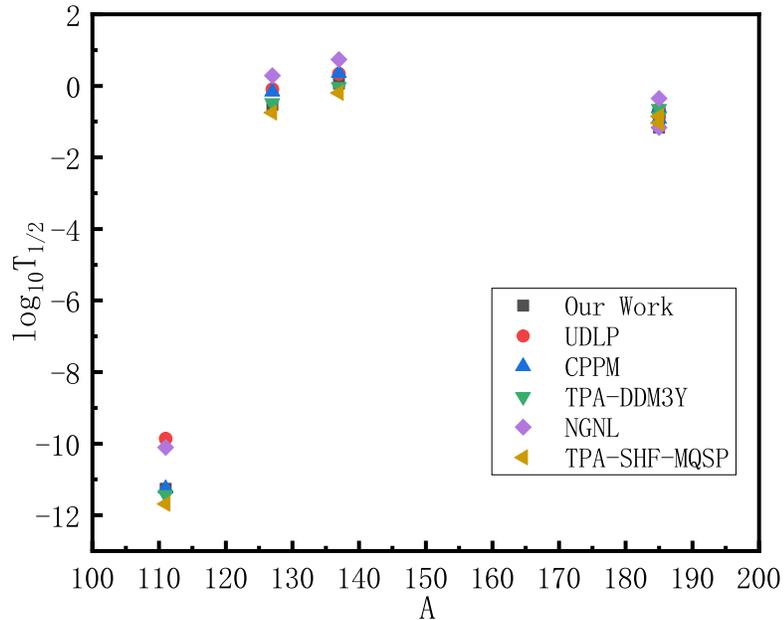}
\caption{The theoretical half-life of deformed proton emission calculated by different models.}
\label{fig 3}
\end{figure}

As seen from Tab. \ref{tab2}, for the same parent nucleus, the theoretical calculations of the proton emission half-life for different models are different due to model dependence, with the significant deviations of the calculations exceeding two orders of magnitude. For a more visual comparison of these theoretical predictions, the theoretical half-lives of proton emission taken from Tab. \ref{tab2} are plotted in Fig. \ref{fig 3}. From this figure, we can clearly see that for all nuclei, the theoretical half-lives of proton emission obtained by the TPA-SHF-MQSP are smaller than other models. On the contrary, the proton emission half-life calculated by the NGNL is more significant than other models. This implies that the proton emission half-life prediction results of TPA-SHF-MQSP and NGNL are potentially under and over-predicted. The prediction results calculated by our model are in the middle of all models, indicating that we work in good agreement with most of the models.

\begin{figure}
\begin{minipage}{0.48\linewidth}
 \centerline{\includegraphics[height=6.0cm,width=7.0cm]{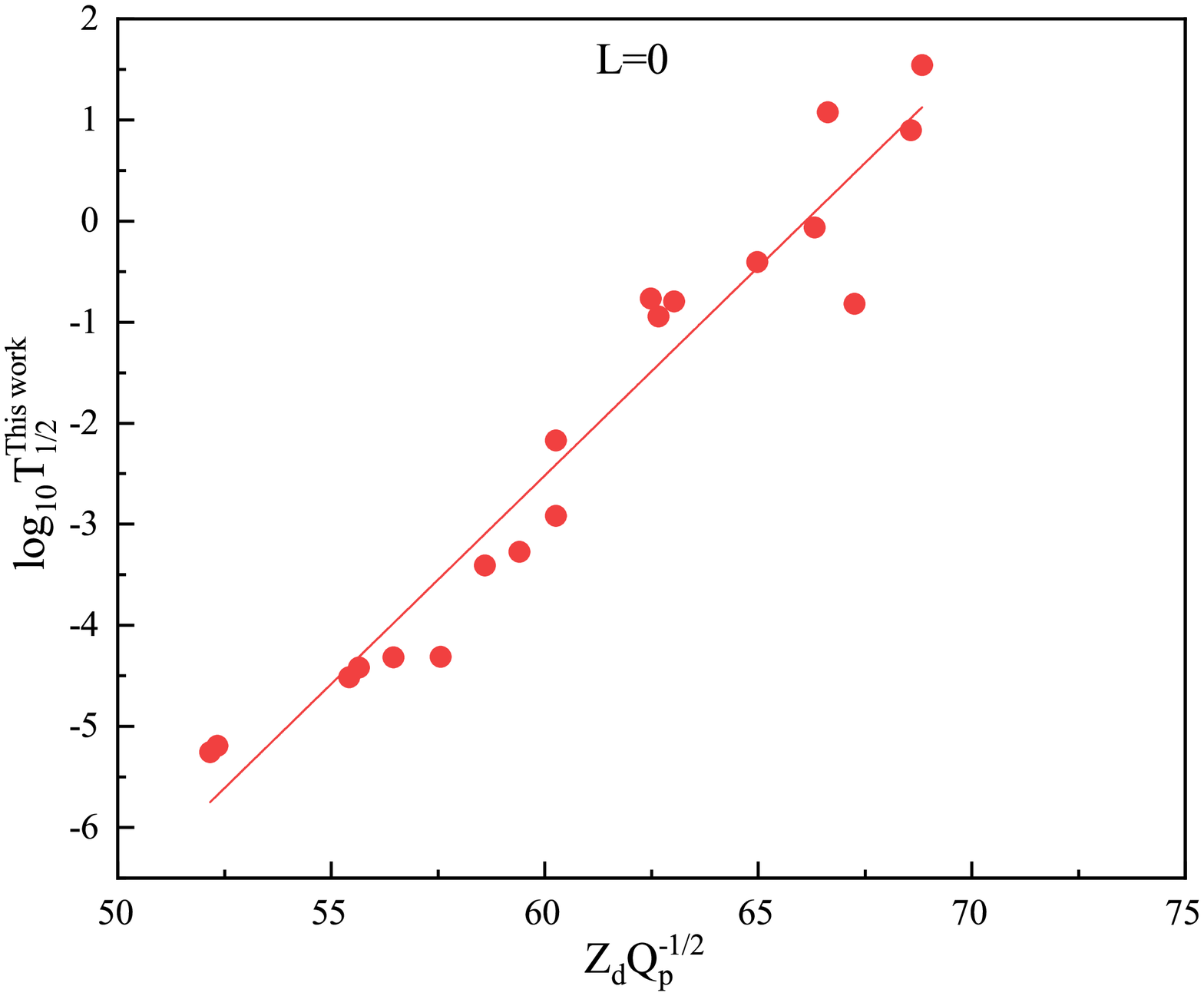}}
 \centerline{(a)}
\end{minipage}
\hfill
\begin{minipage}{0.48\linewidth}
 \centerline{\includegraphics[height=6.0cm,width=7.0cm]{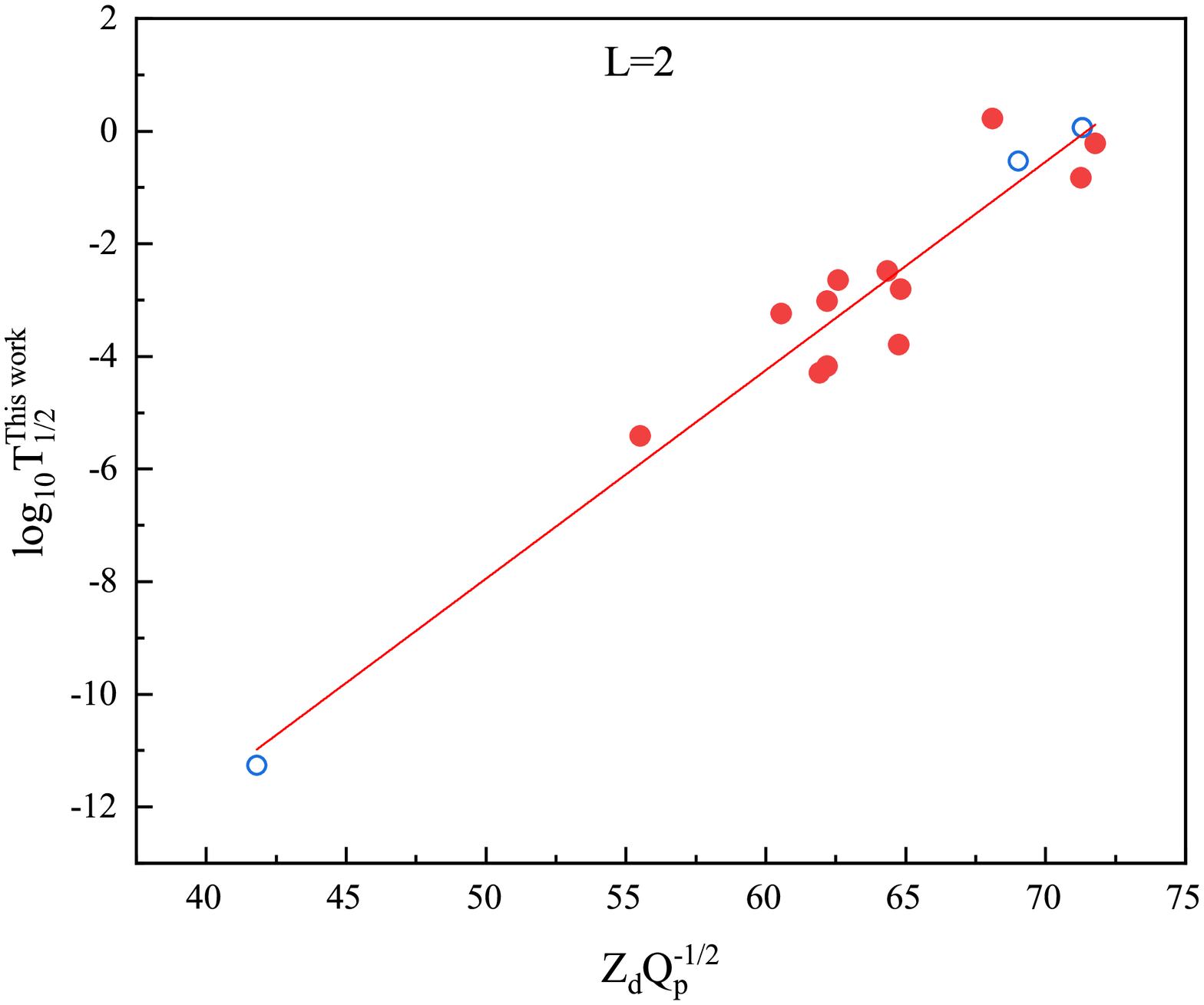}}
 \centerline{(b)}
\end{minipage}
\vfill
\begin{minipage}{.48\linewidth}
 \centerline{\includegraphics[height=6.0cm,width=7.0cm]{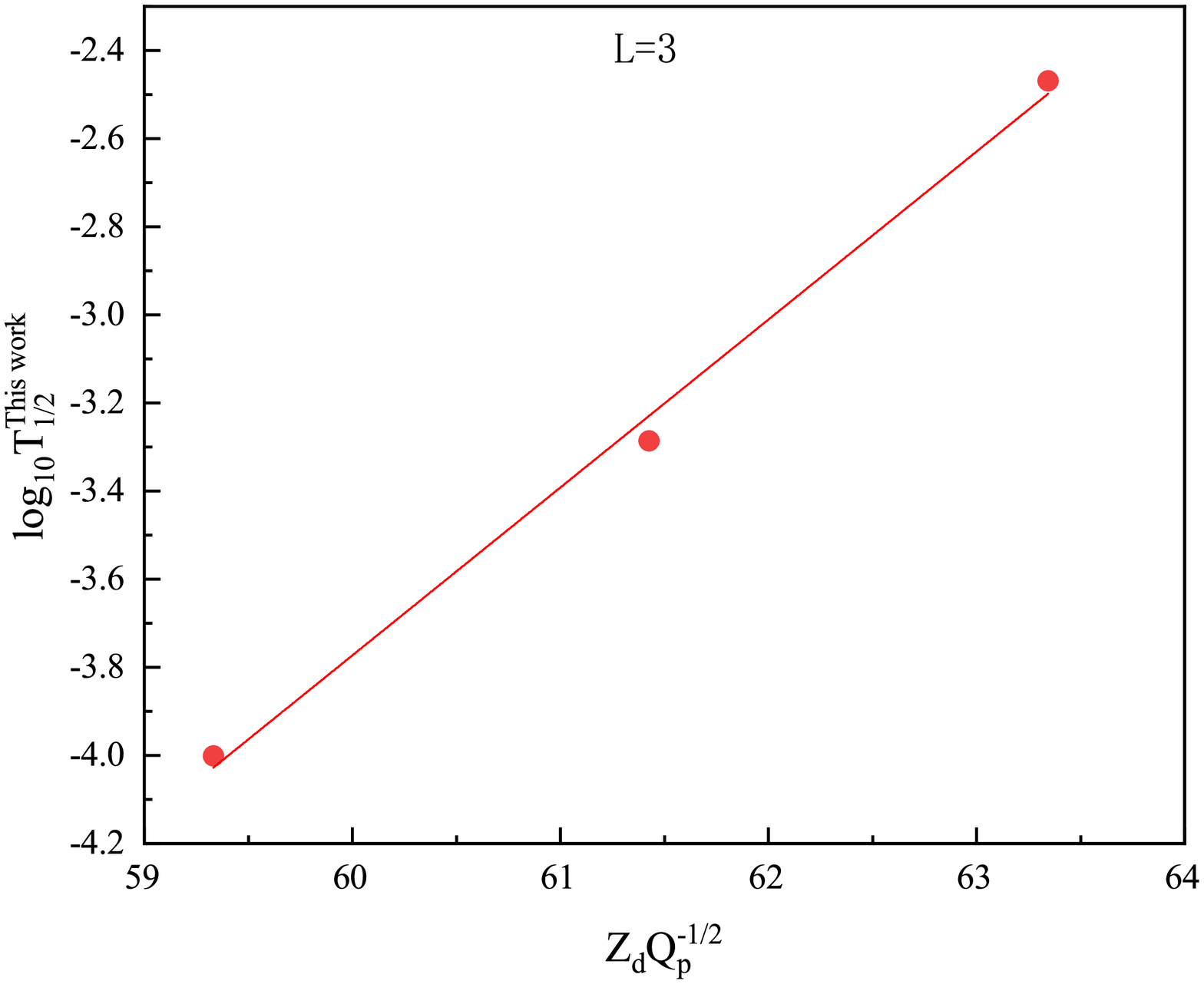}}
 \centerline{(c)}
\end{minipage}
\hfill
\begin{minipage}{0.48\linewidth}
 \centerline{\includegraphics[height=6.0cm,width=7.0cm]{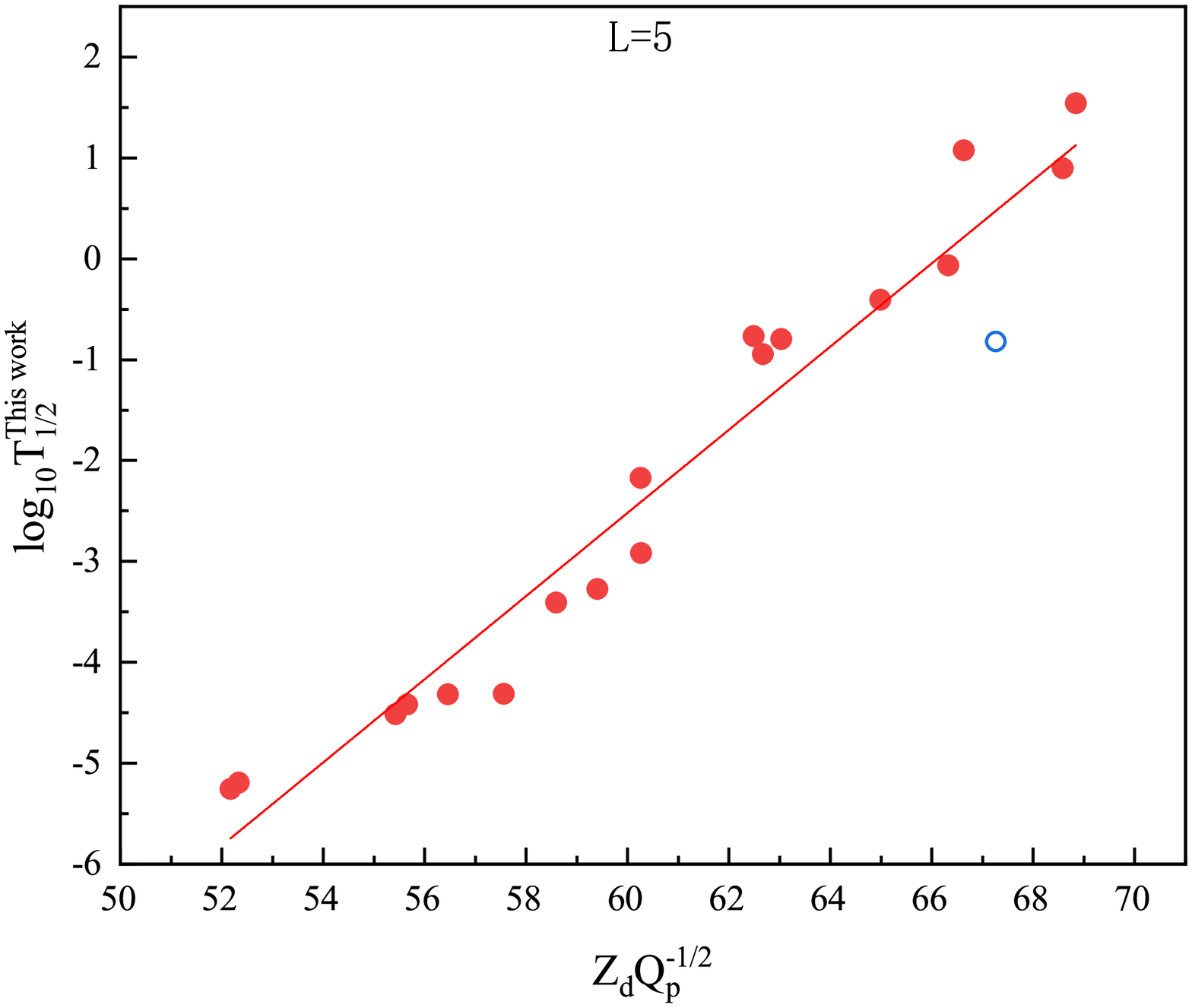}}
 \centerline{(d)}
\end{minipage}
\caption{The linear relationship between ${\rm{lg}{T^{\rm{This-work}}_{1/2}}}$ and $Z_dQ^{-1/2}_p$. The linear relationships are shown in parts (a), (b), (c) and (d) for $l=0, 2, 3$ and 5, respectively. The solid and hollow circles represent the theoretical half-lives calculated by the deformed Gamow-like model, which are taken from Tab. \ref{tab1} and Tab. \ref{tab2}, respectively.}
\label{fig 4}
\end{figure}

The most direct correlation between the half-life and decay energy discovered by Geiger and Nuttall is known as the Geiger-Nuttall law \cite{doi:10.1080/14786441008637156}. By studying the Geiger-Nuttall law, many critical theories were born, e.g., the universal decay law \cite{PhysRevC.92.064301,PhysRevLett.103.072501, PhysRevC.80.044326}, the new Geiger-Nuttall law \cite{PhysRevC.85.044608, epjaChen2019}, Brown-type empirical formula \cite{PhysRevC.46.811} and so on \cite{PhysRevC.85.011303}. Considering the effect of the Coulomb parameters $Z_dQ^{-1/2}_p$ and the orbital angular momentum $l$, the linear relationship between ${\rm{lg}{T^{\rm{This-work}}_{1/2}}}$ and $Z_dQ^{-1/2}_p$ is plotted in Fig. \ref{fig 4}, with $l$ = 0, 2, 3 and 5 labeled as (a), (b), (c) and (d), respectively. This figure shows that ${\rm{lg}{T^{\rm{This-work}}_{1/2}}}$ are linearly dependent on $Z_dQ^{-1/2}_p$ in the case of the orbital angular momentum $l$ kept constant. With the change of the $l$, the slope of $l$ will be affected and changed in succession. Therefore, these linear relationships can also confirm the above statement about the effect of proton emission on orbital angular momentum $l$. Moreover, the well-linear relationships indicate that all the theoretical calculated half-lives taken from Tab. \ref{tab1} and Tab. \ref{tab2} within the deformed Gamow-like model coincide in terms of the Geiger-Nuttall law, which shows our predicted results are reliable.

\subsection{$\mathcal{\alpha}$ decay}

Similarly, the least squares principle was used to fit the adjustable parameters $r_0$ and $h$ of $\mathcal{\alpha}$ decay. $\mathcal{\alpha}$ decay experimental data of 190 even-even nuclei as the database to determine the parameter the radius constant $r_0$, while $h=0$. Then $\mathcal{\alpha}$ decay experimental data of 137 even-$Z$,odd-$N$ nuclei, 124 odd-$Z,$even-$N$ nuclei and 123 doubly-odd nuclei to fit the hindrance factor with the fixed $r_0$. All experimental data are from NUBASE2020 \cite{NUBASE2020}, AME2020 \cite{CPC-2021-0034,CPC-2020-0033} and FRDM2012 \cite{MOLLER20161}. The preformation probability of the $\mathcal{\alpha}$ particle $S_\mathcal{\alpha}=0.5$ remains consistent with Ref.\cite{1402-4896-2013-T154-014029}. The fitting results of 4 adjustable parameters are given by

\begin{equation}
\label{16}
r_0=1.19 {\ } \rm{fm}, \textit{h}_n=0.400, \textit{h}_p=0.346, \textit{h}_{np}=0.528,
\end{equation}
where the parameters $h_{\rm{n}}$, $h_{\rm{p}}$, $h_{\rm{np}}$ are used to describe the effects of an odd-proton, an odd-neutron, an odd-proton and an odd-neutron, respectively.

\begin{figure}[H]
\centering 
\includegraphics[width=14.0cm]{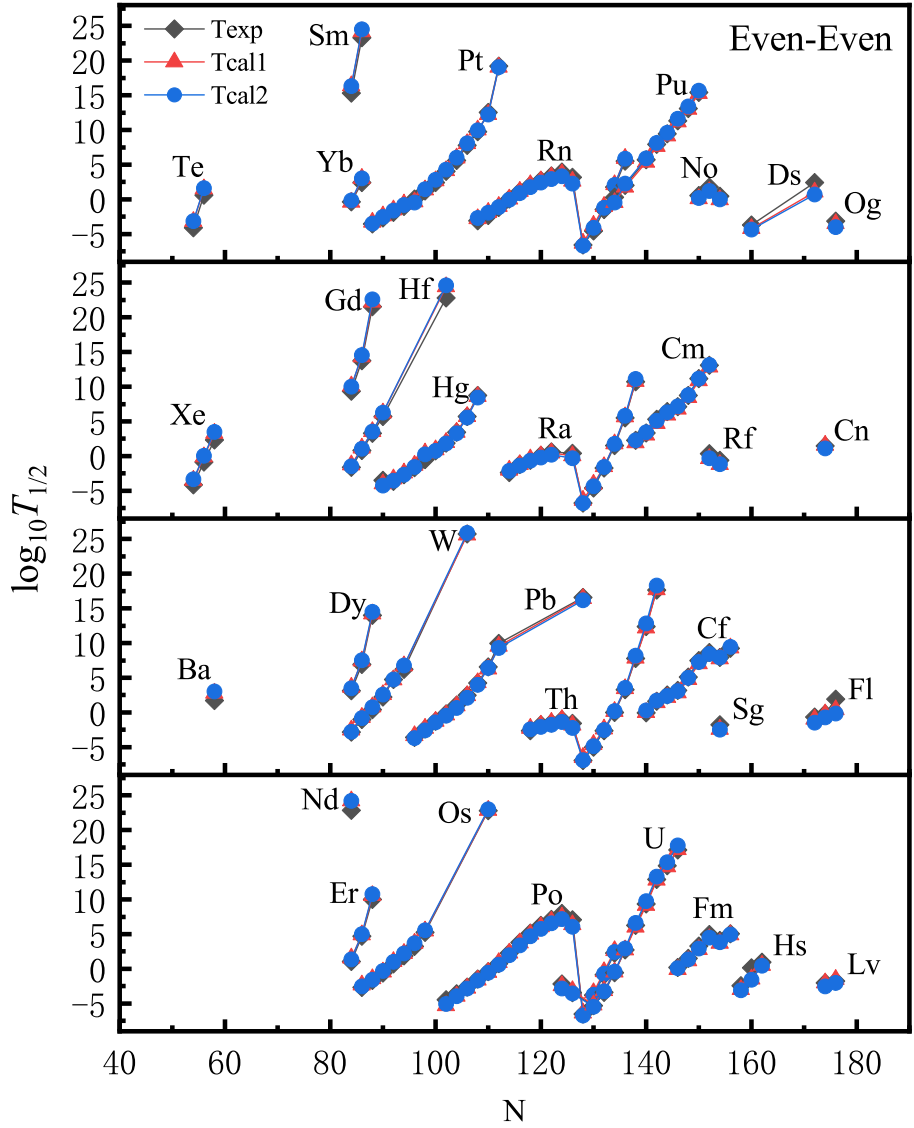}
\caption{The calculations of $\mathcal{\alpha}$ decay half-lives and the experimental data of the even-even nuclei. ${T_{cal1}}$ and ${T_{cal2}}$ are the logarithmic form of $\mathcal{\alpha}$ decay half-life calculated in our work and Gamow-like model, respectively.}
\label{fig 8}
\end{figure}

\begin{figure}[H]
\centering 
\includegraphics[width=14.0cm]{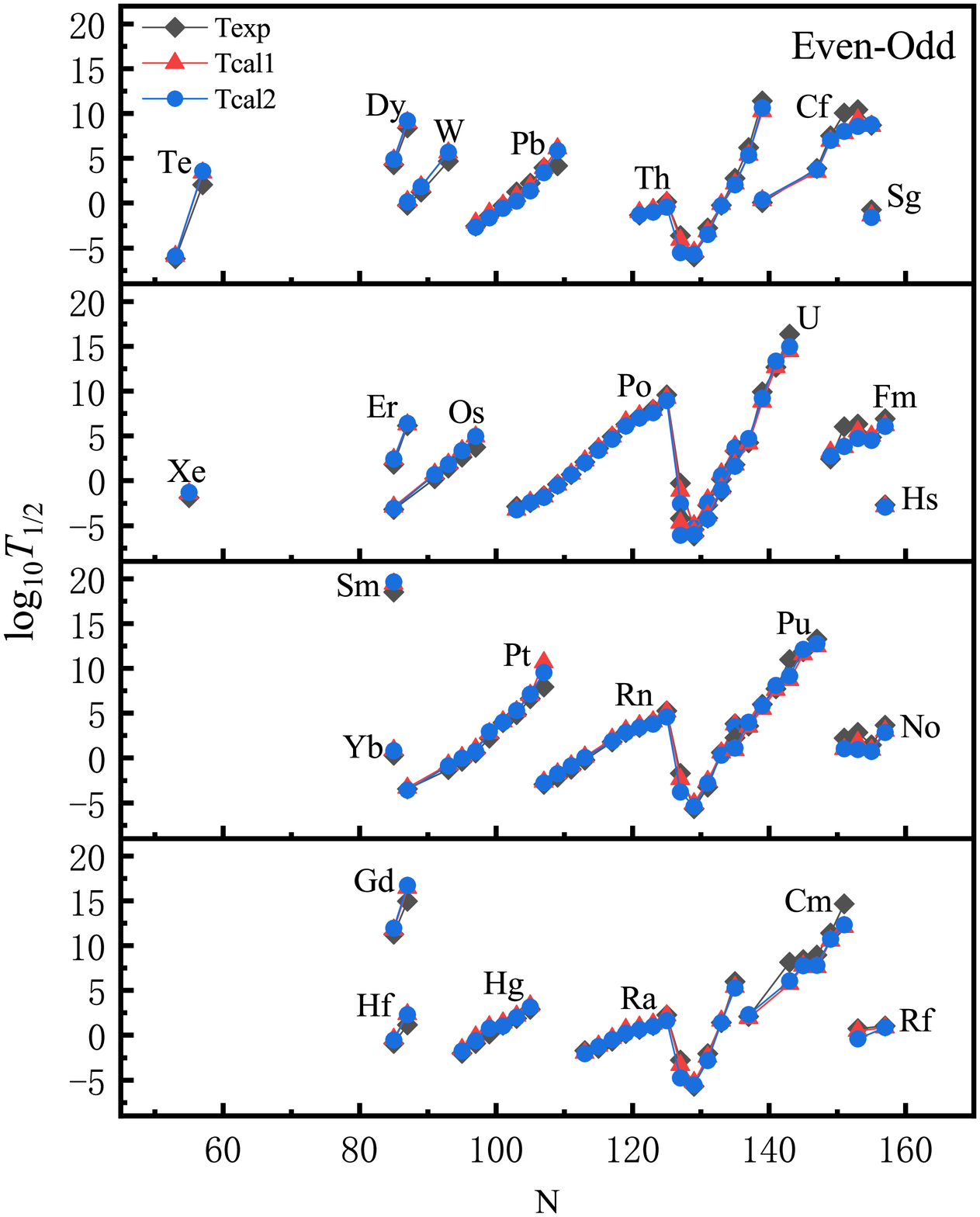}
\caption{The same as Fig. \ref{fig 8}, but for the case of even-$Z$, odd-$N$ nuclei.}
\label{fig 9}
\end{figure}

\begin{figure}[H]
\centering 
\includegraphics[width=14.0cm]{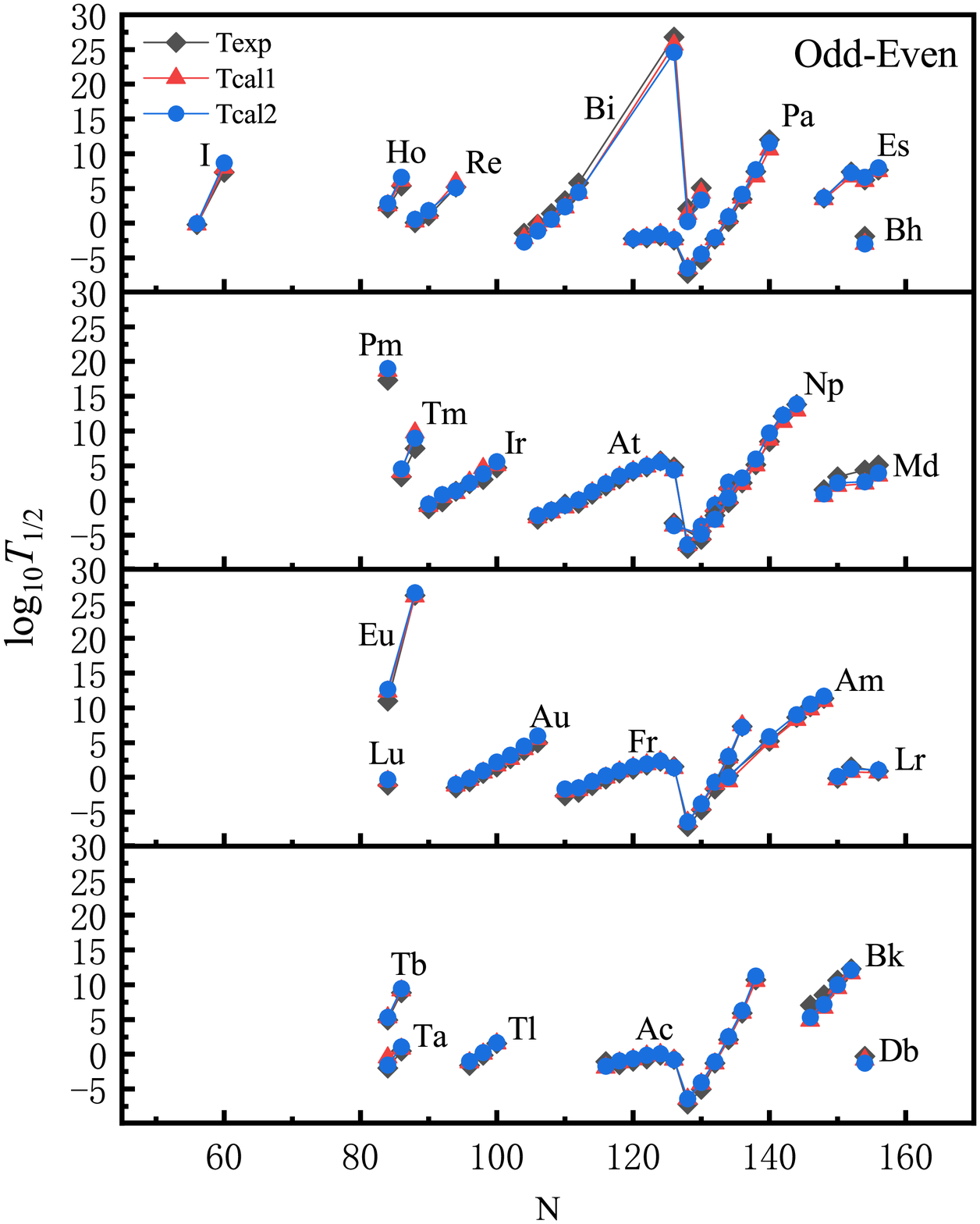}
\caption{The same as Fig. \ref{fig 8}, but for the case of odd-$Z$, even-$N$ nuclei.}
\label{fig 10}
\end{figure}

\begin{figure}[H]
\centering 
\includegraphics[width=14.0cm]{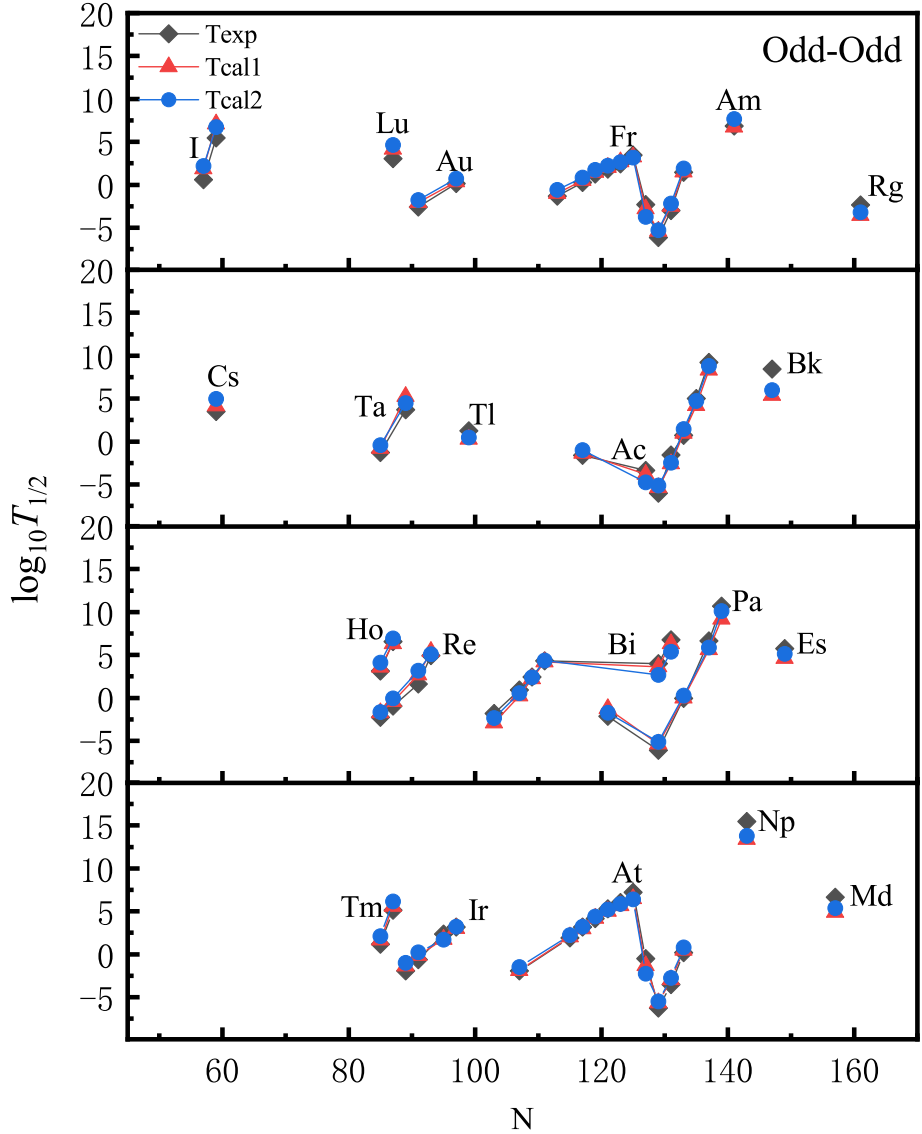}
\caption{The same as Fig. \ref{fig 8}, but for the case of odd-$Z$, odd-$N$ nuclei.}
\label{fig 11}
\end{figure}

In the present work, $\mathcal{\alpha}$ decay half-lives of even-even, odd-$A$, odd-odd nuclei calculated by the Gamow-like model and the deformed Gamow-like model are shown in the Fig. \ref{fig 8} -- \ref{fig 11}. In these figures, the X-axis and the Y-axis represent the neutron number of parent nuclei and the logarithmic form of $\mathcal{\alpha}$ decay half-lives. The black squares represent of the experimental data. The red triangles and the blue circles represen $\mathcal{\alpha}$ decay half-life calculated by our work and Gamow-like model denoted as ${T_{\rm{cal1}}}$ and ${T_{\rm{cal2}}}$, respectively. The cases of even-even nuclei, even-$Z$, odd-$N$ nuclei, odd-$Z$, even-$N$ nuclei and doubly-odd nuclei are shown in Fig. \ref{fig 8}, Fig. \ref{fig 9}, Fig. \ref{fig 10} and Fig. \ref{fig 11}, respectively. 

From this figures, we can found that that our work reproduces the experimental data for most parent nuclei better than Gamow-like model. Moreover, Figs. \ref{fig 8} -- \ref{fig 11} also show that the plunges and peaks in the half-lives, since $N=126$ is magical cores, which plays a crucial role in the $\mathcal{\alpha}$ preformation probability. To quantify the effect of considering the deformation on the calculated $\mathcal{\alpha}$ decay half-lives, we calculated the standard deviation between the theoretical half-lives and the experimental data. The standard deviations between ${T_{\rm{cal1}}}$ and ${T_{\rm{exp}}}$ are denoted as $\sigma_1$, the standard deviations between ${T_{\rm{cal2}}}$ and ${T_{\rm{exp}}}$ are denoted as $\sigma_2$, and the degree of compliance with the experimental data from Figs. \ref{fig 8} -- \ref{fig 11} boosted by considering deformation is denoted as $\sigma_{improve}$, which is given in Tab. \ref{table 3}.

\begin{table}[H]
\centering
\caption{The root-mean-square deviations of the experimental data between Gamow-like model and our work. In the first column, ${\pi_{\rm{z}}}$ and ${\pi_{\rm{n}}}$ are parity of the number of protons and neutrons, respectively. The second column stands for the corresponding total number of nuclei. The following three columns are the hindrance factor $h$ corresponding to the different parity nuclei. The last three columns represent the standard deviations between ${T_{\rm{cal1}}}$ and ${T_{\rm{exp}}}$, the standard deviations between ${T_{\rm{cal2}}}$ and ${T_{\rm{exp}}}$, and the degree of compliance with the experimental data boosted by considering deformation, respectively.}
\label{table 3}

\begin{tabular}{cccccccc}
\multicolumn{1}{c}{${\pi_{\rm{z}}}-{\pi_{\rm{n}}}$}&n&$h_{\rm{n}}$&$h_{\rm{n}}$&$h_{\rm{np}}$&$\sigma_1$&$\sigma_2$&$\sigma_{improve}$\\
\hline
\\
e-e&190&{--}&{--}&{--}&{0.453}&{0.516}&{12.21\%}\\
\\
e-o&137&{0.400}&{--}&{--}&{0.785}&{0.846}&{7.21\%}\\
\\
o-e&124&{--}&{0.346}&{--}&{0.712}&{0.761}&{5.3\%}\\
\\
o-o&123&{--}&{--}&{0.528}&{0.841}&{0.909}&{7.5\%}\\
\hline
\end{tabular}
\end{table}

\begin{figure}[H]
\centering 
\includegraphics[width=16.0cm]{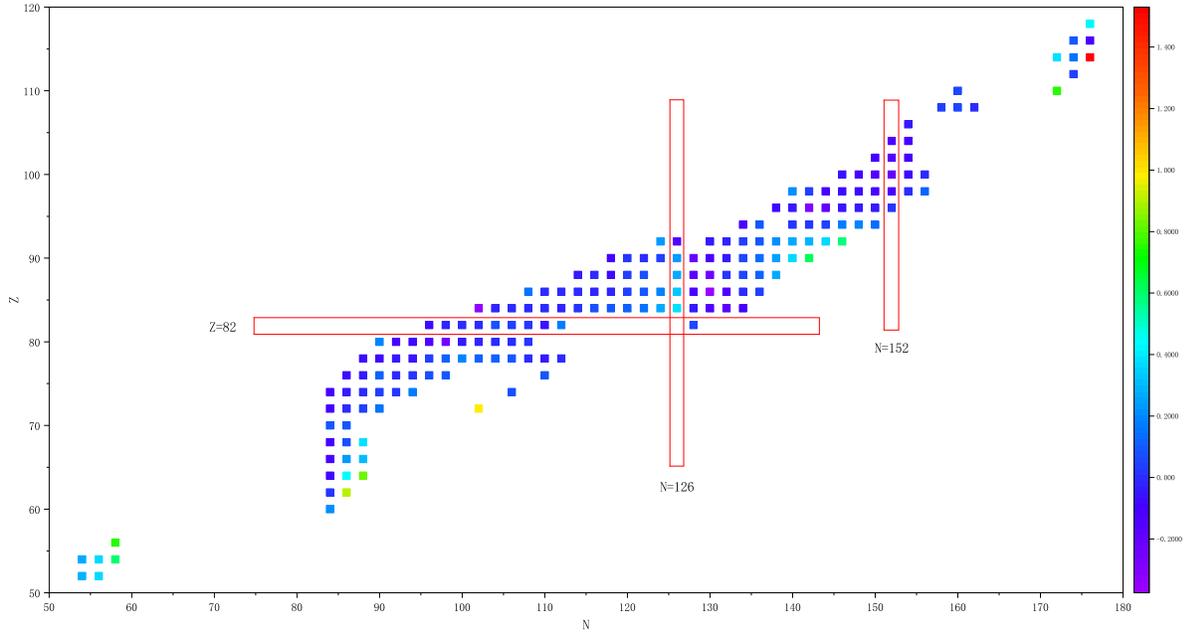}
\caption{The distribution of $\sigma_{s}$ with $N$ and $Z$ of even-even nuclei.}
\label{fig 12}
\end{figure}

From Tab. \ref{table 3}, we can clearly see that for any kinds of the parent nuclei, considering the deformation in the calculations resulting in ${T_{\rm{cal1}}}$ more consistent with the experimental data than ${T_{\rm{cal2}}}$. And the even-even nuclei show the most significant improvement in the degree of conformity with the experimental data. To investigate the influence of the deformation on the calculated half-lives of individual even-even nuclei, we define the enhancement in the degree of conformity of the theoretical half-life of individual even-even nuclei and the experimental data as $\sigma_{s}$,
\begin{equation}
\label{eq5}
\sigma_{s}=({T_{\rm{cal2}}}-{T_{\rm{exp}}})^2-({T_{\rm{cal1}}}-{T_{\rm{exp}}})^2.
\end{equation}
It is easy to find that an immense value of $\sigma_{s}$ represents a better agreement between the theoretical value of $\mathcal{\alpha}$ decay considering the deformation and the experimental data. The distribution of $\sigma_{s}$ with $N$ and $Z$ is given in Fig. \ref{fig 12}. The X-axis and the Y-axis represent the neutron number and the proton number of the parent nucleus, respectively. The starting point of the coordinate axis $N=50$, $Z=50$ and the red box $Z=82$, $N=126$ represent the known magic number.

\begin{figure}[H]
\centering 
\includegraphics[width=13.0cm]{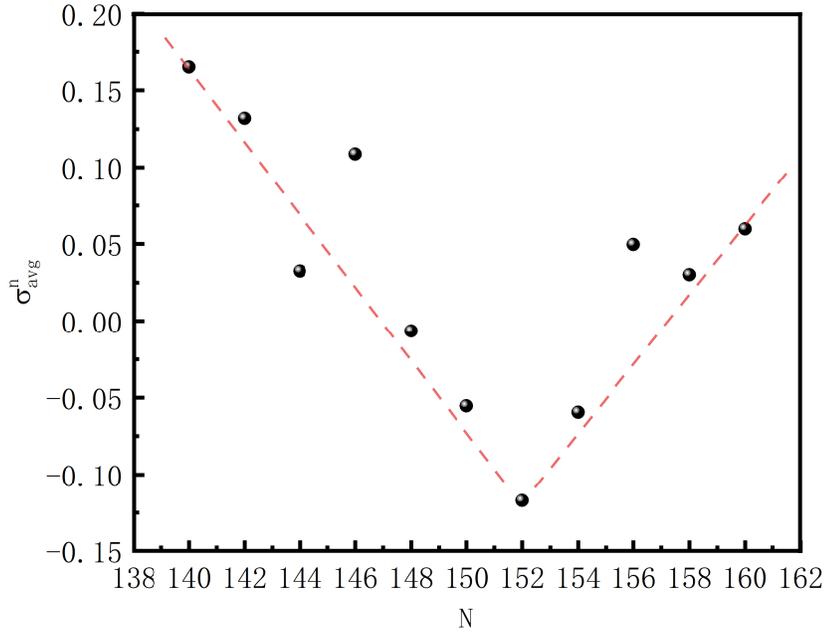}
\caption{The distribution of $\sigma^n_{avg}$ with $N$ and $Z$ of even-even nuclei.}
\label{fig 13}
\end{figure}

Interestingly, the parent nucleus at the shell mostly corresponds to negative values of $\sigma_{s}$. The further away from the shell, the larger the $\sigma_{s}$ value. Nowadays, the research and synthesis of superheavy nuclei (SHN) have become one hot topic in nuclear physics \cite{PhysRevC.97.064609, 1674-1137-41-7-074106, PhysRevC.98.014618}, and finding the number of the next shell is the key to studying superheavy nuclei. Due to the properties of the deformed Gamow-like model concerning the shell structure, we will predict the following neutron magic number in this work. The average value of $\sigma_{s}$ for the same neutron number parent nuclei is defined as $\sigma^n_{avg}$,
\begin{equation}
\sigma^n_{avg}=\sum^{k} \sigma_{s}/k,
\end{equation}
where $k$ represents the number of parent nuclei with the same neutron number. The smaller $\sigma^n_{avg}$ represents the minor effect of considering the deformation on the conformity of $\mathcal{\alpha}$ decay theoretical half-life with the experimental data, and the corresponding $N$ is closer to the magic number. $\sigma^n_{avg}$ with $140 \leq N \leq 160$ is depicted in Fig. \ref{fig 13}. And we can find an excellent regularity in the variation of $\sigma^n_{avg}$ with $N$, which verifies our previous conclusion. Moreover, the value of $\sigma^n_{avg}$ gradually becomes smaller for $140 \leq N \leq 152$ and larger for $152 \leq N \leq 160$, with the lowest value of $\sigma^n_{avg}$ taken at $N = 152$, which indicates a high probability of the following neutron magic number is 152.

\section{Summary}
In summary, we modify Gamow-like model by introducing the nucleus's deformation to study $\mathcal{\alpha}$ decay and proton emission half-lives. The calculation shows that considering deformation for different nuclei with different decay types leads to a better overall agreement of Gamow-like model with the experimental data. Moreover, for nuclei far from the shell, it is necessary to consider the deformation in the calculation. Furthermore, we also used the improved model to predict the proton emission half-lives of the nuclei far from the shell and the number of the next neutron shell. The deformed Gamow-like model shows good agreement with most of other models in predicting proton emission. And we find that number of the next neutron shell is likely to be 152. This work is helpful for future research on proton emission and superheavy nuclei.
\label{section 4}

\section*{Acknowledgements}
This work was supported by the National Key R\&D Program of China (Grant No. 2018YFA0404802), National Natural Science Foundation of China (Grant No. 11875319), the Science and Technology Innovation Program of Hunan Province (Grant No. 2020RC4020), and the Hunan Provincial Innovation Foundation for Postgraduate (Grant No. CX20210007).


\section*{References}

\begin{thebibliography}{}

\bibitem{Oganessian_2015}Oganessian Y T and Utyonkov V K 2015 {\it Rep. Prog. Phys.} \textbf{78} 036301
\bibitem{KARNY200852}Karny M, Rykaczewski K P and Grzywacz R K \emph{et al.} 2008 {\it Phys. Lett. B} \textbf{664} 52
\bibitem{PhysRevC.73.031301}Mohr P 2006 {\it Phys. Rev. C} \textbf{73} 031301
\bibitem{PhysRevC.92.014602}Denisov V Y, Davidovskaya O I and Sedykh I Y 2015 {\it Phys. Rev. C} \textbf{92} 014602
\bibitem{PhysRevC.85.044608}Ren Y J and Ren Z Z 2012 {\it Phys. Rev. C} \textbf{85} 044608
\bibitem{0954-3899-39-1-015105}Poenaru D N, Gherghescu R and Greiner W 2012 {\it J. Phys. G} \textbf{39} 015105
\bibitem{PhysRevC.80.024310}Delion D S 2002 {\it Phys. Rev. C} \textbf{80} 024310
\bibitem{0954-3899-42-5-055112}Wang Z, Niu Z, Liu Q and Guo J 2015 {\it J. Phys. G} \textbf{42} 055112
\bibitem{1674-1137-41-1-014102}Sun X D, Wu X J, Zheng B, Xiang D, Guo P and Li X H 2017 {\it Chin. Phys. C} \textbf{41} 014102
\bibitem{PhysRevC.94.024338}Sun X D, Guo P and Li X H 2016 {\it Phys. Rev. C} \textbf{94} 024338
\bibitem{PhysRevC.93.034316}Sun X D, Guo P and Li X H 2016 {\it Phys. Rev. C} \textbf{93} 034316
\bibitem{PhysRevC.95.014319}Sun X D, Duan C, Deng J G, Guo P and Li X H 2017 {\it Phys. Rev. C} \textbf{95} 014319
\bibitem{PhysRevC.95.044303}Sun X D, Duan C, Deng J G, Xiang D, Guo P and Li X H 2017 {\it Phys. Rev. C} \textbf{95} 044303
\bibitem{PhysRevC.96.024318}Deng J G, Zhao J C, Xiang D and Li X H 2017 {\it Phys. Rev. C} \textbf{96} 024318
\bibitem{PhysRevC.97.044322}Deng J G, Zhao J C, Chu P C and Li X H 2018 {\it Phys. Rev. C} \textbf{97} 044322
\bibitem{PhysRevLett.65.2975}Buck B, Merchant A C and Perez S M 1990 {\it Phys. Rev. Lett.} \textbf{65} 2975
\bibitem{PhysRevC.74.014304}Xu C and Ren Z Z 2006 {\it Phys. Rev. C} \textbf{74} 014304
\bibitem{XU2005303}Xu C and Ren Z Z 2005 {\it Nucl. Phys. A} \textbf{760} 303
\bibitem{0305-4616-5-10-005}Poenaru D N, Ivascu M and Sandulescu A 1979 {\it J. Phys. G} \textbf{5} L169
\bibitem{PhysRevC.48.2409}Goncalves M and Duarte S B 1993 {\it Phys. Rev. C} \textbf{48} 2409
\bibitem{0954-3899-26-8-305}Royer G 2000 {\it J. Phys. G} \textbf{26} 1149
\bibitem{PhysRevC.74.017304}Zhang H F, Zuo W, Li J Q and Royer G 2006 {\it Phys. Rev. C} \textbf{74} 017304
\bibitem{GUO2015110}Guo S Q, Bao X J, Gao Y, Li J Q and Zhang H F 2015 {\it Nucl. Phys. A} \textbf{934} 110
\bibitem{PhysRevLett.59.262}Gurvitz S A and Kalbermann G 1987 {\it Phys. Rev. Lett. } \textbf{59} 262
\bibitem{SANTHOSH201528}Santhosh K P, Sukumaran I and Priyanka B 2015 {\it Nucl. Phys. A} \textbf{935} 28
\bibitem{PhysRevC.87.024308}Zdeb A, Warda M and Pomorski K 2013 {\it Phys. Rev. C} \textbf{87} 024308
\bibitem{0954-3899-31-2-005}Tavares O, Medeiros E and Terranova M L 2005 {\it J. Phys. G} \textbf{31} 129
\bibitem{PhysRevC.81.064318}Ni D D and Ren Z Z 2010 {\it Phys. Rev. C} \textbf{81} 064318
\bibitem{QI2014203}Qi C, Andreyev A N, Huyse M, Liotta R J, Duppen P V and Wyss R 2014 {\it Phys. Lett. B} \textbf{734} 203
\bibitem{PhysRevC.72.051601}Basu D N, Chowdhury P R and Samanta C 2005 {\it Phys. Rev. C} \textbf{72} 051601
\bibitem{Qian_2010}Qian Y B, Ren Z Z, Ni D D and Sheng Z Q 2010 {\it Chin. Phys. Lett.} \textbf{27} 112301
\bibitem{PhysRevC.96.034619}Santhosh K P and Sukumaran I 2017 {\it Phys. Rev. C} \textbf{96} 034619
\bibitem{PhysRevC.56.1762}\AA{}berg S and Semmes P B and Nazarewicz W 1997 {\it Phys. Rev. C} \textbf{56} 1762
\bibitem{PhysRevC.85.011303}Qi C, Delion D S, Liotta R J and Wyss R 2012 {\it Phys. Rev. C} \textbf{85} 011303
\bibitem{FERREIRA2011508}Ferreira L S and Maglione E and Ring P 2011 {\it Phys. Lett. B} \textbf{701} 508
\bibitem{PhysRevC.79.054330}Dong J M, Zhang H F and Royer G 2009 {\it Phys. Rev. C} \textbf{79} 054330
\bibitem{Zhang_2010}Zhang H F, Wang Y J, Dong J M, Li J Q and Scheid W 2010 {\it J. Phys. G} \textbf{37} 085107
\bibitem{PhysRevC.95.014302}Wang Y Z, Cui J P, Zhang Y L, Zhang S and Gu J Z 2017 {\it Phys. Rev. C} \textbf{95} 014302
\bibitem{PhysRevC.71.014603}Balasubramaniam M and Arunachalam N 2005 {\it Phys. Rev. C} \textbf{71} 014603
\bibitem{BHATTACHARYA2007263}Bhattacharya M and Gangopadhyay G 2007 {\it Phys. Lett. B} \textbf{651} 263
\bibitem{Chen_2019}Chen J L, Li X H, Cheng J H, Deng J G and Wu X J 2019 {\it J. Phys. G} \textbf{46} 065107
\bibitem{1402-4896-2013-T154-014029}Zdeb A, Warda M and Pomorski K 2013 {\it Phys. Scr.} \textbf{2013} 014603
\bibitem{Liu_2021}Liu H M, Pan X, Zou Y T, Chen J L, Cheng J H, He B and Li X H 2021 {\it Chin. Phys. C} \textbf{45} 044110
\bibitem{CHENG2019350}Cheng J H, Chen J L, Deng J G, Wu X J, Li X H and Chu P C 2019 {\it Nucl. Phys. A} \textbf{987} 350
\bibitem{Liu_2022}Liu H M, Zou Y T, Pan X, Li X H, Wu X J and He B 2021 {\it Phys. Scr.} \textbf{96} 125322
\bibitem{Zhu2022}Zhu D X, Xu Y Y, Liu H M, Wu X J, He B and Li X H 2022 {\it Eur. Phys. J. A} \textbf{33} 122
\bibitem{XING2022122528}Xing F Z, Qi H, Liu H M, Cui J P, Gao Y H, Wang Y Z, Gu J Z and Yong G C 2022 {\it Nucl. Phys. A} \textbf{1028} 122528
\bibitem{Azeez2022}Azeez O K, Yahya W A and Saeed A A 2022 {\it Phys. Scr.} \textbf{97} 055302
\bibitem{Zdeb2016}Zdeb A, Warda M, Petrache C M and Pomorski K 2016 {\it Eur. Phys. J. A} \textbf{52} 323
\bibitem{PhysRevC.83.014601}Poenaru D N, Gherghescu R A and Greiner W 2011 {\it Phys. Rev. C} \textbf{83} 014601
\bibitem{PhysRevC.81.064309}Dong J M, Zuo W, Gu J Z, Wang Y Z and Peng B B 2010 {\it Phys. Rev. C} \textbf{81} 064309
\bibitem{PhysRevC.69.024614}Xu C and Ren Z Z 2004 {\it Phys. Rev. C} \textbf{69} 024614
\bibitem{Gur31}Morehead J J 1995 {\it J. Math. Phys.} \textbf{36} 5431
\bibitem{PhysRevC.82.059901}Denisov V Y and Khudenko A A 2009 {\it Phys. Rev. C} \textbf{82} 059901
\bibitem{MOLLER20161}Möller P, Sierk A J, Ichikawa T and Sagawa H 2016 {\it Atom. Data Nucl. Data} \textbf{109} 1
\bibitem{PhysRevC.73.041301}Xu C and Ren Z Z 2006 {\it Phys. Rev. C} \textbf{73} 041301
\bibitem{PhysRevC.61.044607}Takigawa N, Rumin T and Ihara N 2000 {\it Phys. Rev. C} \textbf{61} 044607
\bibitem{ISMAIL200353}Ismail M, Seif W M and El-Gebaly H 2003 {\it Phys. Lett. B} \textbf{563} 53
\bibitem{Gao_Long_2008}Zhang G L, Le X Y and Liu Z H 2008 {\it Chin. Phys. Lett.} \textbf{25} 1247
\bibitem{CPC-2021-0034}Huang W J, Wang M, Kondev F G, Audi G and Naimi S 2021 {\it Chin. Phys. C} \textbf{45} 030002
\bibitem{CPC-2020-0033}Wang M, Huang W J, Kondev F G, Audi G and Naimi S 2021 {\it Chin. Phys. C} \textbf{45} 030003
\bibitem{NUBASE2020}Kondev F G, Wang M, Huang W J, Naimi S and Audi G 2021 {\it Chin. Phys. C} \textbf{45} 030001
\bibitem{BLANK2008403}Blank B and Borge M J G 2008 {\it Prog. Part. Nucl. Phys.} \textbf{60} 403
\bibitem{PhysRevLett.103.072501}Qi C, Xu F R, Liotta R J and Wyss R 2009 {\it Phys. Rev. Lett.} \textbf{103} 072501
\bibitem{PhysRevC.80.044326}Qi C, Xu F R, Liotta R J, Wyss R and Zhang M Y 2009 {\it Phys. Rev. C} \textbf{80} 044326
\bibitem{PhysRevC.93.024612}Ghodsi O N and Daei-Ataollah A 2016 {\it Phys. Rev. C} \textbf{93} 024612
\bibitem{GUO201354}Guo C L, Zhang G L and Le X Y 2013 {\it Nucl. Phys. A} \textbf{897} 54
\bibitem{Deng2019}Deng J G, Li X H, Chen J L, Cheng J H and Wu X J 2019 {\it Eur. Phys. J. A} \textbf{55} 58
\bibitem{Chen2018}Chen J L, Cheng J H, Deng J G and Li X H 2018 {\it Nucl.
Phys. Rev.} \textbf{35} 257
\bibitem{epjaChen2019}Chen J L, Xu J Y, Deng J G, Li X H, He B and Chu P C 2019 {\it Eur. Phys. J. A} \textbf{55} 214
\bibitem{Cheng_2022}Cheng J H, Zhang Z, Wu X J, Chu P C and Li X H 2022 {\it Chin. Phys. C} \textbf{46} 104104
\bibitem{doi:10.1080/14786441008637156}Geiger H P D and Nuttall J M B S 1911 {\it Philos. Mag.} \textbf{22} 613
\bibitem{PhysRevC.92.064301}Wang Y Z, Wang S J, Hou Z Y and Gu J Z 2015 {\it Phys. Rev. C} \textbf{95} 064301
\bibitem{PhysRevC.46.811}Brown B A 1992 {\it Phys. Rev. C} \textbf{46} 811
\bibitem{PhysRevC.97.064609}Wu Z H, Zhu L, Li F, Yu X B, Su J and Guo C C 2018 {\it Phys. Rev. C} \textbf{97} 064609
\bibitem{1674-1137-41-7-074106}Liu J H, Guo S Q, Bao X J and Zhang H F 2017 {\it Chin. Phys. C} \textbf{41} 074106
\bibitem{PhysRevC.98.014618}Li F, Zhu L, Wu Z H, Yu X B, Su J and Guo C C 2018 {\it Phys. Rev. C} \textbf{98} 014618









\end{thebibliography}
\end{document}